\documentclass[10pt,conference]{IEEEtran}
\IEEEoverridecommandlockouts
\usepackage{cite}
\usepackage{amsmath,amssymb,amsfonts}
\usepackage{graphicx}
\usepackage{textcomp}
\usepackage{hyperref}
\usepackage{xcolor}
\usepackage{xspace}
\usepackage{color}
\usepackage[normalem]{ulem}
\definecolor{TUGreen}{rgb}{0.517,0.721,0.094}
\definecolor{TUOrange}{rgb}{1.0,0.7176,0.0}
\definecolor{BrightGray}{gray}{0.9}
\definecolor{DarkGray}{gray}{0.2}
\definecolor{white}{rgb}{1,1,1}
\definecolor{black}{rgb}{0,0,0}
\definecolor{red}{rgb}{1,0,0}

\usepackage{listings}
\definecolor{codegreen}{rgb}{0,0.6,0}
\definecolor{codegray}{rgb}{0.5,0.5,0.5}
\definecolor{codepurple}{rgb}{0.58,0,0.82}
\definecolor{backcolour}{rgb}{0.95,0.95,0.92}
\definecolor{lstBg}{gray}{0.95}

\lstdefinestyle{mylst}{
	breaklines=true,
	breakindent=10pt,
	escapeinside={?}{?},
	numbers=left,
	basicstyle=\footnotesize,
	numbersep=2pt,
	tabsize=2, %
	backgroundcolor=\color{lstBg},
	captionpos=b,
	frame=none,
	columns=fullflexible, %
	belowcaptionskip=-2mm,
	escapeinside={(*@}{@*)} 
}
\lstdefinelanguage{realACSL}[]{C}{
	deletecomment=[s]{/*}{*/},
	deletecomment=[l]//,
	keywordsprefix=\\,
	keywords = [2]{assumes, requires, ensures, behavior, assigns, calls, logic, predicate,disjoint,complete,behaviors,ghost,global,invariant,assert},
	keywordstyle=[2]\color{ForestGreen},
}

\def\BibTeX{{\rm B\kern-.05em{\sc i\kern-.025em b}\kern-.08em
		T\kern-.1667em\lower.7ex\hbox{E}\kern-.125emX}}

\usepackage{booktabs}  
\usepackage{etoolbox}

\providebool{techreport}
\setbool{techreport}{false}

\usepackage{array}
\usepackage{graphicx}
\usepackage{amsmath}

\usepackage{setspace}
\usepackage{algorithm}
\usepackage{algorithmic}
\usepackage{tikz}
\usepackage{xcolor}
\usepackage{multirow}
\usepackage{paralist}
\usepackage{url}
\usepackage{etoolbox}
\usepackage{cancel}
\usepackage{tabu}
\usepackage{comment}
\usepackage{mathtools}
\usepackage{marvosym}

\usetikzlibrary{calc,shadows,patterns,shapes,arrows,decorations.pathmorphing,backgrounds,positioning,fit,plotmarks}
\tikzset{>=latex}
\usetikzlibrary{shapes,arrows}
\usetikzlibrary{backgrounds,patterns}
\usetikzlibrary{positioning,fit,calc}
\tikzstyle{decision} = [diamond,aspect=2, draw, fill=white, 
text width=5em, text badly centered, node distance=2.15cm, inner sep=0pt]
\tikzstyle{block} = [rectangle, draw, fill=white, 
text width=14em, text centered, rounded corners, minimum height=2em, node distance=1.25cm]
\tikzstyle{line} = [draw,->]
\tikzstyle{cloud} = [draw, ellipse,fill=red!20, 
minimum height=2em]

\usepackage{amsthm}

\newtheorem{definition}{Definition}

\newtheoremstyle{propertystyle}
{0pt} %
{0pt} %
{}    %
{}    %
{\bfseries} %
{:}   %
{.3em} %
{}    %

\theoremstyle{propertystyle}

\newtheorem{property}{Property}

\def\myendproof{{\ \vbox{\hrule\hbox{%
				\vrule height1.3ex\hskip0.8ex\vrule}\hrule }}\par}

\usepackage{array}
\newcolumntype{L}[1]{>{\raggedright\let\newline\\\arraybackslash\hspace{0pt}}m{#1}}
\newcolumntype{C}[1]{>{\centering\let\newline\\\arraybackslash\hspace{0pt}}m{#1}}
\newcolumntype{R}[1]{>{\raggedleft\let\newline\\\arraybackslash\hspace{0pt}}m{#1}}

\tikzset{
	task/.style={shade, shading=radial, rectangle,minimum height=.1cm,
		inner color=#1!20, outer color=#1!60!gray},
	task1/.style={task=yellow, minimum width=13mm},
	task2/.style={task=orange, minimum width=13mm},
	task3/.style={task=red, minimum width=13mm},
	task4/.style={task=green, minimum width=13mm},
	task5/.style={task=blue, minimum width=13mm},
	task6/.style={task=purple, minimum width=13mm},
	task7/.style={task=cyan, minimum width=13mm},
	task8/.style={task=pink, minimum width=13mm},
}

\floatname{algorithm}{Algorithm}

\newif\ifshowcomments
\showcommentstrue
\ifshowcomments
  \newcommand{\jj}[1]{\textcolor{red}{\emph{jj(}} \textcolor{blue}{#1} \textcolor{red}{)jj}}
  \newcommand{\junjie}[1]{\textcolor{blue}{junjie: #1 : endjunjie \\}}
  \newcommand{\kuan}[1]{\textcolor{magenta}{kuan: #1}}
  \newcommand{\kay}[1]{\textcolor{cyan}{kay: #1}}
  \newcommand{\jan}[1]{\textcolor{green!60!black}{jan: #1}}
  \newcommand{\dk}[1]{\textcolor{orange}{dk: #1}}
  \newcommand{\todo}[1]{\textcolor{red}{\textsf{TO DO: #1}}}
\else
  \newcommand{\jj}[1]{}
  \newcommand{\junjie}[1]{}
  \newcommand{\kuan}[1]{}
  \newcommand{\kay}[1]{}
  \newcommand{\jan}[1]{}
  \newcommand{\dk}[1]{}
  \newcommand{\todo}[1]{}
\fi

\begin{document}

	\title{Deductive Verification for Earliest Deadline First Scheduler Implementations}

	\author{%
		\IEEEauthorblockN{Daniel Kuhse\IEEEauthorrefmark{1},
			Junjie Shi\IEEEauthorrefmark{1},
			Jan Duy Thien Pham\IEEEauthorrefmark{1},
			Kay Heider\IEEEauthorrefmark{1},
			Marcus V\"olker\IEEEauthorrefmark{2},
			Kuan-Hsun Chen\IEEEauthorrefmark{3},
			Jian-Jia Chen\IEEEauthorrefmark{1}\IEEEauthorrefmark{2}}
		\IEEEauthorblockA{\IEEEauthorrefmark{1}TU Dortmund University, Germany\\
			\{daniel.kuhse, junjie.shi, jan.pham, kay.heider\}@tu-dortmund.de}
		\IEEEauthorblockA{\IEEEauthorrefmark{2}RWTH Aachen University, Germany\\
			voelker@embedded.rwth-aachen.de, jian-jia.chen@rwth-aachen.de}
		\IEEEauthorblockA{\IEEEauthorrefmark{3}University of Twente, The Netherlands\\
			k.h.chen@utwente.nl}
	}

	\thispagestyle{plain}
	
	\maketitle
	\IEEEpeerreviewmaketitle
	
	\begin{abstract}
		
		Real-Time Operating Systems (RTOSes) rely on scheduler implementations to provide predictable task execution. 
		For safety-critical systems, it is therefore not sufficient to reason only about the abstract scheduling policy; the concrete implementation must also preserve the intended scheduling semantics. 
		This is particularly challenging for Earliest Deadline First (EDF) scheduling, because EDF introduces dynamic, deadline-derived priorities that are often realized by reusing kernel infrastructure originally designed for fixed-priority scheduling.
			
		In this work, we formalize EDF correctness through three essential properties that any implementation of the Earliest Deadline First (EDF) scheduler must satisfy. Based on these properties,	we propose a framework utilizing deductive verification, that applies to any EDF-based scheduler realization. We instantiate the framework in Frama-C/ACSL and apply it to three structurally different EDF scheduler realizations: RTEMS 5, RTEMS 6, and an EDF extension of FreeRTOS. 
		The verification confirms that the considered implementations satisfy the EDF correctness properties under explicitly stated assumptions on kernel infrastructure. 
		
	\end{abstract}
	

\section{Introduction}
\label{sec:introduction}

Real-time systems play an indispensable role in modern computing, where timely and predictable responses are crucial across various domains, such as aerospace, automotive, medical devices, and industrial control. In these contexts, Real-Time Operating Systems (RTOSes) act as the backbone, managing hardware resources and ensuring that real-time tasks are executed within their specified deadlines.

Ensuring scheduler correctness requires more than proving the abstract scheduling theory. 
Even when a scheduling policy is well understood mathematically, the corresponding implementation may deviate from the intended semantics because of legacy APIs, implementation optimizations, or interactions with existing kernel data structures. 
This is particularly relevant for widely used open-source RTOSes such as FreeRTOS~\cite{freertos_web}, Zephyr~\cite{zephyr_web} and RTEMS~\cite{rtems_web}, . Unlike clean-slate kernels such as seL4~\cite{klein2009sel4} and CertiKOS~\cite{DBLP:conf/osdi/GuSCWKSC16} which are designed from the ground up for formal verification, they are developed and maintained over long periods and often evolve by extending existing kernel infrastructure. Formal verification of scheduler implementations in these existing RTOS is therefore important, but difficult: the verification must account for implementation-level data structures and function interactions while remaining modular enough to be applicable to real systems.

This work focuses on Earliest Deadline First (EDF) scheduling. 
EDF has been extensively studied and is optimal for preemptive uniprocessor scheduling under classical assumptions.
Unlike fixed-priority scheduling, however, EDF assigns priorities dynamically according to absolute deadlines: jobs with earlier deadlines must receive higher scheduling priority. 
This creates additional implementation complexity. 
A job release may require recomputing a deadline-derived priority, propagating this value through internal priority structures, repositioning the job in the ready queue, and triggering a preemption decision. 
If any of these steps is missed or performed in the wrong order, the implementation may violate EDF semantics even though the high-level scheduling policy is conceptually simple.

In practice, EDF support is often built on top of kernel infrastructure that was not originally designed for dynamic-priority scheduling. 
For example, RTEMS realizes EDF through its existing priority-management and scheduler-node infrastructure.
FreeRTOS, by contrast, does not provide EDF scheduling by default, so an EDF scheduler must be introduced by modifying its fixed-priority scheduling infrastructure. 
These implementations illustrate a common challenge: EDF correctness is not confined to the task-selection policy alone, but depends on the consistent maintenance of deadline-derived priorities across task releases, ready-queue updates, and scheduler decisions.

Implementation optimizations further amplify this problem. For example, Zephyr uses \texttt{\_kernel.ready\_q.cache} to cache the next job to execute, avoiding repeated ready-queue searches. 
As reported in~\cite{zephyr_bug}, incorrect maintenance of this cache can lead to unintended execution orders. Specifically, due to this design choice, each job must update its absolute deadline before being added to the ready queue, as no further deadline updates are permitted once the job is in the queue unless the \texttt{\_kernel.ready\_q.cache} is explicitly refreshed. 
Such cases show that documentation and programming discipline alone are insufficient: EDF correctness must be verified across the interactions between deadline updates, ready-queue state, and scheduler decisions. At the same time, comprehensive verification of existing RTOS kernels is difficult. 
Current verification tools must cope with legacy APIs, low-level data structures, and implementation-specific control flow. 
A modular verification strategy is therefore commonly used: individual components are verified under explicit assumptions about the correctness of the surrounding kernel infrastructure.

Existing verification efforts address related aspects, but do not fully cover this problem. 
Some approaches develop verified kernels or scheduler components within verification-oriented frameworks, such as seL4~\cite{klein2009sel4} and CertiKOS~\cite{DBLP:conf/osdi/GuSCWKSC16}. Vanhems et al.~\cite{DBLP:conf/rtas/VanhemsRNG22} proposed a formal proof methodology for EDF scheduling by reimplementing the scheduler, rather than directly verifying the existing implementations.

Other approaches focus on existing RTOSes, but verify selected APIs, abstract scheduler models, or applications running on top of RTOS semantics. For example, Liang et al.~\cite{liang2016correctness} verify 22 API functions in the FreeRTOS scheduling module.  However, it lacks a systematic framework for reasoning about the correct composition of these functions.
The deductive framework proposed in~\cite{DBLP:journals/tecs/TascheHH25} targets FreeRTOS applications under cooperative (non-preemptive) scheduling. 
It relies on a formal encoding of FreeRTOS scheduling semantics and therefore verifies application behavior with respect to this encoding.
However, these works do not verify how EDF scheduling is realized inside the scheduler infrastructure of existing RTOS implementations. 
This is important because EDF correctness is not confined to the task-selection policy alone: it also depends on how deadline-derived priorities are represented, propagated, and maintained across scheduler nodes, ready queues, and task release operations.

\noindent\textbf{Our Contribution:}
In this work, we propose a general framework for verifying the correctness of EDF scheduler implementations in existing RTOSes.
To the best of our knowledge, this is the first work to present a formal verification framework applicable to EDF scheduler implementations in existing RTOSes. The main challenges lie in identifying correctness properties, formalizing them within a general framework, and addressing the reuse of generic kernel primitives not originally designed for dynamic-priority scheduling.
The contributions of this work are summarized as follows:
\begin{itemize}
	
\item We analyze the general operational workflow of EDF schedulers in RTOSes and formally define three key correctness properties, \emph{deadline priorities}, \emph{ready queue management}, and \emph{scheduler decisions}, that any correct implementation must satisfy.

\item We propose a deductive verification framework for formally verifying EDF scheduler implementations against the \textit{three} defined correctness properties under clearly stated assumptions. 

\item We validate the framework across various EDF scheduler realizations, including RTEMS 5 and RTEMS 6~(Sec.~\ref{sec:verification_edf}). We also modify a standard FreeRTOS V11.1.0 to support EDF scheduling and subsequently verify it~(Sec.~\ref{sec:freertos-edf}).
This demonstrates portability across evolving and structurally distinct RTOS infrastructures. We make our verification effort including the ACSL annotations, FreeRTOS modification and replication instructions available as an artifact~\cite{git_source_edf_verif}.\footnote{\url{https://github.com/TU-Dortmund-CS-LS12-DAES-teaching/edf-verification}}
\end{itemize}

\section{Background}
\label{sec:background}
In this section, we first introduce the task model assumed in this work. We then describe the fundamental principles of the EDF scheduling algorithm. Finally, we provide a formal definition of the key properties that any correct EDF scheduler implementation must satisfy.
\subsection{Periodic Task Model}
\label{sec:system-model}
We consider the implemented scheduler is employed on a uni-processor system for a set of $N$ tasks \mbox{$\mathbb{T} = \{\tau_1,~\ldots,~\tau_N\}$}.
Each task is described by \mbox{$\tau_i=(D_i, T_i)$}, %
where:
\begin{itemize}
	
	\item $T_i$ is the period of task $\tau_i$.%
	
	\item $D_i$ is the relative deadline of  $\tau_i$.%
\end{itemize}
All tasks release an infinite number of task instances, called jobs. %
Each job is characterized by a tuple \mbox{$J_i^\ell=(\tau_i, r_i^\ell, d_i^\ell)$}. For the $\ell$-th job, $\ell \in \mathbb{N}$: $r_i^\ell$ is the release time, i.e., $r_i^\ell = \ell \times T_i$,
and $d_i^\ell$ is the absolute deadline, i.e., $d_i^\ell = r_i^\ell + D_i$.
Please note that the worst-case execution time (WCET) $C_i$ of task $\tau_i$ is not relevant to the EDF scheduler's operation and is thus omitted from the system model.

Let the ready queue $\textbf{Q}(t) = \{J_1, J_2, \dots, J_n\}$ be a list of jobs sorted in non-decreasing order of their absolute deadlines. %
Any two jobs $J_i^\imath, J_j^\jmath \in \textbf{Q}(t)$ 
must satisfy the following condition: if $J_i^\imath$ precedes $J_j^\jmath$ in $\textbf{Q}(t)$, then $d_i^\imath \leq d_j^\jmath$. If $d_i^\imath = d_j^\jmath$, then $J_i^\imath$ must be released earlier than $J_j^\jmath$.
In addition, the schedule function $\textbf{S}: \mathbb{N} \rightarrow J \cup \{ \perp \}$ maps each time $t$ to the job $J$ executing at time $t$, or $\perp$ if the processor is idle.

\subsection{EDF Scheduling Workflow and Core Components}
\label{sec:general-concept}

\begin{figure}
	\centering
	\includegraphics[width=\linewidth]{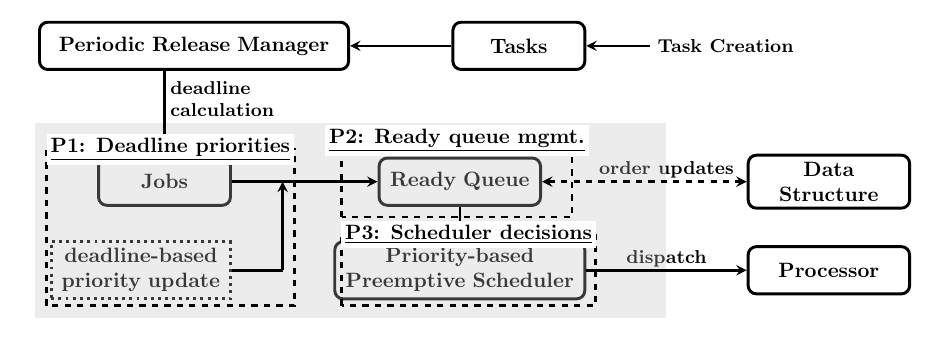}
	\caption{The workflow of EDF scheduler. The gray area is the verification target in this work. Each block marks the area relevant to a property.}
	\label{fig:rtems-edf-flow}
\end{figure}

Preemptive EDF scheduling %
guarantees that, at any moment, the job with the earliest absolute deadline among all ready jobs is selected for execution.
Figure~\ref{fig:rtems-edf-flow} illustrates the general operational workflow of an EDF scheduler, abstracted across RTOS implementations. The process begins with task creation. %
The Periodic Release Manager handles task activations by generating jobs at each release point, computing absolute deadlines, and assigning priorities accordingly.
Afterwards, released jobs are inserted into the ready queue %
with deadline-based ordering. The Priority-Based Preemptive Scheduler monitors the queue and the currently executing job, performing preemptions when a new job arrives with an earlier deadline.

While the core scheduling principle is well-defined, its realization in RTOS kernels can vary significantly due to legacy design choices, internal API constraints, and system-level optimizations. Despite these differences, most EDF implementations share three core functionalities, including:

\begin{itemize}
	\item \uline{Priority assignment}: EDF is a job-level dynamic-priority scheduler that assigns priorities to jobs of real-time tasks based on their absolute deadlines, i.e., the earlier the deadline, the higher the priority. When adopting RTOS routines designed for fixed-priority scheduling, priorities of real-time tasks are calculated and updated whenever a new job arrives.
	\item \uline{Ready queue management}: The ready queue holds all ready jobs ordered by the deadline-derived priorities established above. Whenever a job's priority is assigned or changes, the queue is updated accordingly, so that the head is consistently the job with the earliest absolute deadline in the ready queue.
	\item \uline{Preemptive EDF scheduling}: On job arrivals, completions, or priority changes, the scheduler selects the head of the ready queue, i.e., the job with the earliest deadline, for execution. EDF operates in a preemptive manner. That is, if a newly released job has an earlier deadline than the currently running preemptible job, it immediately preempts the executing job. This triggers a context switch.
\end{itemize}

When multiple jobs in the ready queue share the same priority (i.e., identical absolute deadlines), the scheduling order is implementation-defined. For example, in the RTEMS implementation, tie-breaking is resolved using FIFO semantics, i.e., the job enqueued earlier is given the higher priority.%

\subsection{Formal Properties Required for EDF Scheduler}
\label{sec:rtems-edf-properties}

We formalize the correctness of EDF scheduling as a single property over the externally observable state, and introduce a set of supporting subproperties that together establish it.

\begin{definition}\label{def:edf-correctness}
    \textbf{Earliest Deadline First.}
    A preemptive EDF scheduler implementation is correct if, at every observable time point $t$ outside the scheduler, the currently running job $J_r$ implies that either 1) $J_r$ has the earliest absolute deadline among the set of eligible jobs $\textbf{Q}(t) \cup \{J_r\}$, or 2) a context switch has been scheduled and will execute before the next observable point outside the scheduler.
\end{definition}

To establish Definition~\ref{def:edf-correctness}, we identify three properties that any correct EDF implementation must satisfy:

\begin{property}\label{prop:deadline-priorities}
    \textbf{Deadline priorities (P1).} For every job, its absolute deadline is mapped to its scheduling priority.
\end{property}

\begin{property}\label{prop:ready-queue}
    \textbf{Ready queue management (P2).} The ready queue is updated according to scheduling events to reflect the set of ready jobs with priorities as established by Property~\ref{prop:deadline-priorities}, and yields its minimum-priority element correctly.
\end{property}

\begin{property}\label{prop:scheduler-decisions}
    \textbf{Scheduler decisions (P3).} Whenever the ready queue is modified, the scheduler selects $\arg\min_{J \in \textbf{Q}(t)} d_J$ from the ready queue. If the running job is preemptible and some ready job has an earlier deadline, a context switch is scheduled.
\end{property}

All of these properties are local to the scheduler. We assume that the context switch is executed immediately after the scheduler finishes the selection of the highest-priority job\footnote{This property holds in FreeRTOS because the context switch is called immediately after the scheduler. For RTEMS, the context switch is deferred until the thread dispatch disable level returns to zero.}. If this assumption does not hold, the implementation results in a significant gap between the selection of the highest-priority job and the actual context switch. This requires an additional verification to ensure that the ready queue is not updated between the end of the scheduler and the start of the subsequent context switch. To establish this would require covering all interrupt handlers and OS infrastructure. This is a broader verification effort related to the correctness of the scheduler of the RTOS itself in addition to its EDF variant, and thus out of scope for this work.

Combined these properties establish Definition~\ref{def:edf-correctness}. Property~\ref{prop:deadline-priorities} and Property~\ref{prop:ready-queue} together ensure that the ready queue includes all ready jobs with their priorities assigned to their deadlines and the head being the job with the earliest deadline. Property~\ref{prop:scheduler-decisions} ensures that after returning from any scheduler entry point the running job has the earliest deadline or a context switch is scheduled. Under the previously discussed assumption for the context switch execution, it is then guaranteed that on exiting the interrupt routine, the running job has the earliest deadline.

\section{General EDF Scheduling Structures}
\label{sec:general-imp-and-verif}

\label{sec:main-components}
This section outlines the key components required for EDF scheduling: the task model, periodic release manager, jobs, ready queue, and priority-based preemptive scheduler. %

\subsubsection{Tasks}

A task represents a long-lived software entity in an RTOS, typically defined at system configuration time. Tasks encapsulate application-level control flows and may be periodic, sporadic, or aperiodic. In EDF scheduling, tasks themselves do not carry static priorities; rather, they act as templates for generating jobs, which are assigned dynamic priorities based on their absolute deadlines.

Each task is associated with a set of configuration parameters that determine its timing behavior, i.e., \mbox{$\tau_i=(D_i, T_i)$}. %
These configurations are typically specified during system initialization but may be updated during runtime to accommodate dynamic system behavior. Each task is represented internally using a Task Control Block (TCB), which stores configuration data and runtime information such as next release time and pointers to pending jobs.

\subsubsection{Periodic Release Manager}

The Periodic Release Manager (PR Manager) is responsible for generating jobs according to the temporal behavior specified by each task, including periodic, sporadic, or event-driven releases. It enforces the timing constraints of tasks and ensures that job creation aligns with the system's real-time requirements.

For each periodic task, the PR Manager maintains a release schedule, typically driven by a system timer. 
When a release condition is met, the PR Manager generates a new job instance, computes its absolute deadline, derives the corresponding EDF priority, and inserts the job into the ready queue.

\subsubsection{Jobs}
A job is a runtime instantiation of a task, characterized by a specific arrival time and absolute deadline. During execution, jobs transition between scheduler-visible states such as Ready, Executing, and Blocked, depending on dispatch decisions and synchronization events.

From the scheduler's perspective, each job release triggers a sequence of ready-queue updates, dispatch decisions, and potential preemption events that determine whether EDF ordering semantics are preserved. For example, a running job may be preempted and returned to the Ready state if another job with an earlier absolute deadline becomes runnable.

In EDF scheduling, a job’s priority is determined by its absolute deadline, with earlier deadlines corresponding to higher priority (i.e., lower numerical values in typical RTOS encodings). Each job maintains scheduler-relevant attributes including its arrival time, absolute deadline, dynamic priority, and links required for ready-queue management.

\subsubsection{Ready Queue}

The ready queue holds all jobs that are ready to execute. Independent of the underlying implementation (e.g., red-black trees or sorted linked lists), EDF correctness requires that queue ordering consistently reflects deadline-derived priorities.

The ready queue provides the following core operations:
\begin{itemize}
	\item Inserting a newly released or preempted job with its computed priority;
	\item Updating a job’s position when its deadline changes;%
	\item Retrieving the job with the earliest deadline (the head of the queue) for scheduling decisions.
		
\end{itemize}

The correctness and efficiency of the ready queue are critical to ensuring timely and predictable scheduling behavior.

\subsubsection{Priority-Based Preemptive Scheduler}

The priority-based preemptive scheduler selects and dispatches the job with the highest priority (i.e., earliest deadline) for execution. It continuously monitors job arrivals, completions, and priority changes to determine whether a context switch is required.

Key operations include:
\begin{itemize}
	\item Comparing the priority of a newly released or updated job with that of the currently running job;
	\item Initiating preemption when a new job has a higher priority, triggering a context switch;
	\item Dispatching the job with the earliest deadline from the ready queue when the processor is idle or when the current job completes;
	\item Performing context switches: saving the state of the current job, enqueuing it back into the ready queue, and restoring the context of the next job.
	
\end{itemize}

The scheduler ensures that the job with the earliest deadline always gains processor access, faithfully enforcing EDF scheduling semantics.

\section{Formal Verification Framework for EDF Scheduler}
\label{sec:formal-verif-edf-general}
This section presents a verification framework targeting the core components of EDF scheduler implementations in RTOSes using Frama-C. In contrast to the approach of~\cite{DBLP:conf/rtas/VanhemsRNG22}, which verifies a reimplemented EDF scheduler built on newly designed primitives, our goal is to formally verify existing implementations embedded within RTOS kernels, potentially relying on generic infrastructure not originally intended for EDF scheduling.

\subsection{Deductive Verification}
\label{sec:deductive-verification}

In order to verify the properties of the EDF scheduler implementation, we apply deductive verification~\cite{DBLP:journals/cacm/Hoare69, floyd_flowcharts_1967}. For completeness, we briefly review Hoare Logic and ACSL.

Hoare Logic provides a systematic framework for rigorously reasoning about program correctness~\cite{DBLP:journals/cacm/Hoare69}. A key concept is the \textit{Hoare Triples}, denoted as $\{P\} \; C \; \{Q\}$, where:
\begin{itemize}
	\item \textit{C} is the set of considered instructions in the program.
	\item \textit{P} is the \textit{precondition}, describing the state before executing \textit{C}.
	\item \textit{Q} is the \textit{postcondition}, describing the expected state after \textit{C}'s execution.
\end{itemize}

If the precondition \textit{P} holds before execution, and \textit{C} terminates, the postcondition \textit{Q} must hold afterward.

To express these properties, we use the ANSI/ISO C Specification Language (ACSL)~\cite{ACSL}, which enables the formal specification of function contracts directly in C code. Deductive verification with function contracts is modular: when verifying a function, other called functions are treated as black boxes and only their contracts are considered, unless explicitly inlined. Contracts are embedded using comments (\texttt{/*@ ... */}) and include:
\begin{itemize}
	\item \texttt{requires} for preconditions;
	\item \texttt{assigns} to declare which variables may be modified;
	\item \texttt{ensures} for postconditions;
	\item \texttt{$\backslash$valid} to specify valid memory regions.
\end{itemize}

ACSL supports \textit{predicates} which describe properties of the program state, which can be used as pre- and postconditions. For example, a predicate could state that an item is in the ready queue or that the head of the ready queue has the earliest deadline. These predicates can be used to specify intended behavior and express invariants. This way we can formally capture the properties outlined in Section~\ref{sec:rtems-edf-properties} as ACSL contracts, which are then verified against the implementation.

\subsection{Frama-C and Memory Model}
\label{sec:framac-and-mm}

To validate the function contracts, we use Frama-C (Framework for Modular Analysis of C programs) version 32~\cite{cuoq2012frama, framaCAbout}, along with the WP plugin~\cite{wpPlugin} for static analysis and deductive verification. The WP plugin generates proof obligations, i.e., logical formulas derived from the code and its annotations, which are initially simplified using the built-in Qed engine. If Qed cannot fully resolve them, the obligations are passed to an SMT solver via Why3~\cite{articlewhy3}. In this work, we use Alt-Ergo version 2.6.3~\cite{alt_ergo_web} as the backend solver. Verification is performed at the source-code level, making the approach architecture-independent, assuming a conforming implementation of lower-level system components.

Accurate verification of C code requires a formal memory model to represent read and write operations, particularly when accessing memory via pointers. We apply the \textit{Typed} memory model~\cite{garion2022gentle}, which is the standard recommendation and represents memory using global arrays and models pointers as indices. The \texttt{\textbackslash separated} predicate allows specifying that two pointers refer to disjoint memory regions, preventing aliasing issues and unintended interactions between components.
We also enable \texttt{Cast} option to support unsafe pointer casts, which are often used in system-level code to implement low-level abstractions and structural inheritance, necessitating it for verifying RTEMS and FreeRTOS.

Some scheduler decisions are represented through volatile state, such as pending-dispatch flags or current-task pointers. 
By default, Frama-C treats each volatile access as potentially yielding a fresh value, which prevents stable reasoning about such variables. 
To verify Property~\ref{prop:scheduler-decisions}, we thus use Frama-C's \texttt{Volatile} option to mirror volatile accesses into proof-only ghost variables. 
The soundness of this approach relies on the absence of concurrent modifications during scheduler operations, which is ensured for the uniprocessor scheduler entry points considered in our case studies.

\subsection{Verification Boundary and Assumptions}
\label{sec:key-assumptions}
Our framework verifies the EDF scheduler against the properties defined in Section~\ref{sec:rtems-edf-properties}, treating certain parts of the RTOS infrastructure as correct components. The modularity of contract-based verification allows us to focus on the core EDF logic. Data structures used by the scheduler are specified at the verification boundary through contracts. These contracts are assumptions, but a verified replacement could be substituted without affecting the EDF proofs. The remaining OS infrastructure is assumed correct implicitly.

\textbf{Correctness of OS infrastructure}: We assume the correctness of underlying kernel primitives. This includes the dispatch mechanism, memory management, context switching and the periodic tick infrastructure. None of these are specific to EDF. For context switching, we additionally assume that a requested context switch is executed before the next observable point outside the scheduler, with no ready-queue modifications intervening, as discussed in Section~\ref{sec:rtems-edf-properties}.

\textbf{Correctness of operations on advanced data structures}: 
We treat scheduler data structures as part of the verification boundary and model each ready queue abstractly as a set of priority nodes. 
Membership predicates and interface contracts specify insertion, removal, and minimum-element retrieval, allowing the EDF proof to rely on abstract ready-queue behavior rather than the concrete data-structure implementation. 
In the case studies below, this abstraction is instantiated for RTEMS, whose ready queue is implemented using OpenBSD's red-black tree, and FreeRTOS, whose ready queue is implemented using a doubly-linked list. 
Additional scheduler lists, such as FreeRTOS delayed/suspended lists and RTEMS priority-aggregation trees, are modeled in the same way. 
Stronger guarantees could be obtained in future work by verifying these data structures or adopting verified alternatives such as priority search trees~\cite{DBLP:journals/afp/LammichN19a}.

\subsection{Verification Strategy}
\label{sec:verif-strategy}
We establish properties~\ref{prop:deadline-priorities}--\ref{prop:scheduler-decisions} through two proof artifacts: 1) predicates to state invariants over the scheduler's state, and 2) function contracts to describe the expected behavior of the scheduler's core components. Key invariants are maintained across the scheduler's operations. We give a high-level overview of the modeling strategy. Our concrete ACSL specifications are very large as they need to account for implementation details, so we opt for idealized versions to clearly illustrate the general pattern, with the full specifications available in the artifact~\cite{git_source_edf_verif}.

The ready queue is modeled as a set of priority nodes 
with predicates describing its state. For example, \texttt{In(j, Q)} expresses 
that job~\texttt{j} is a member of the ready queue~\texttt{Q}, and 
\texttt{head\_is\_earliest(Q)} expresses that the minimum-priority element 
of~\texttt{Q} corresponds to the job with the earliest absolute deadline.

Every scheduler entry point that mutates the ready queue carries postconditions stating which predicates change and how. Listing~\ref{lst:pattern-contract} illustrates the pattern on a generic unblock operation: a queue well-formedness invariant is maintained, the scheduler decision predicate is re-established: either the running job remains earliest or a dispatch is pending, and the postcondition states that the only change to the queue's membership is the addition of the unblocked job~\texttt{j}.

\begin{lstlisting}[language=C,style=mylst,float=tb,caption={The transition contract on a generic unblock entry point.},captionpos=b,label={lst:pattern-contract}, belowskip=-8pt]
/*@
  requires ready_queue_invariant(Q);
  requires edf_running_earliest(Q);
  ...
  ensures ready_queue_invariant(Q);
  ensures edf_running_earliest(Q);
  ensures \forall Job *j'; 
  				In(j', Q) <==> \old(In(j', Q)) || j' == j;
*/
void scheduler_unblock(Queue *Q, Job *j);
\end{lstlisting}

We apply this pattern to establish each property as follows.

\emph{Property~\ref{prop:deadline-priorities}.} An invariant links the scheduler's effective priority back to the job's deadline, paired with a postcondition on the function that originally sets the deadline.

\emph{Property~\ref{prop:ready-queue}.} Ready-queue management needs to maintain an invariant that the head of the queue corresponds to the job with the earliest deadline. Operations that modify the queue have postconditions that reflect the expected changes to the queue's membership. Moreover, the ready queue must be well-formed, such as correct links between nodes and their TCB.

\emph{Property~\ref{prop:scheduler-decisions}.} Scheduler decisions are 
captured by a single predicate stating that either the running job has the 
earliest deadline among all ready jobs, or a dispatch is pending, see Listing~\ref{lst:edf-running-earliest}. This predicate is carried as a postcondition on every scheduler entry point.
\begin{lstlisting}[language=C,style=mylst,float=tb,caption={Scheduler predicate for the ready queue, global running job.},captionpos=b,label={lst:edf-running-earliest},belowskip=-15pt]
/*@ predicate edf_running_earliest(Queue *Q) =
      dispatch_scheduled ||
      (\forall Job *j; In(j, Q) ==>
         running_job->deadline <= j->deadline);
*/
\end{lstlisting}

\section{Verification of the EDF scheduler in RTEMS}
\label{sec:verification_edf}

Through the proposed framework, we conduct a case study on the EDF scheduler implementation in RTEMS 5 and 6, focusing on the scheduler module and relevant \texttt{\_Thread} functions. We give an overview of the EDF implementation, outline the preprocessing steps, summarize key EDF-related functions, and explain their verification objectives.

\subsection{EDF Implementation Overview}

The RTEMS scheduling infrastructure is built around a base scheduler node that stores job-specific metadata required for scheduling decisions. These decisions are triggered by events such as job completions and ready-queue updates. A base node includes a priority aggregation, which maintains a red-black tree of all active priority nodes associated with it. A priority node captures an individual priority contribution to the thread, for example from the scheduler itself, an EDF deadline, or synchronization-related priority inheritance. This allows RTEMS to combine priority contributions from different sources, such as EDF deadlines and synchronization-related priority adjustments.

\begin{figure}[t]
	\centering
	\includegraphics[width=\linewidth]{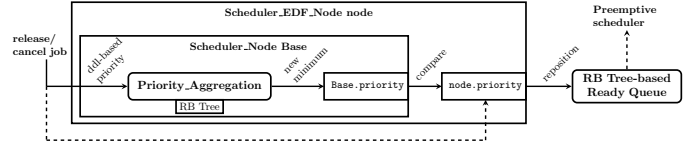}
	\caption{The ready queue management workflow within EDF scheduler node in RTEMS implementation.}
	\label{fig:priority-levels}
\end{figure}

Each scheduling algorithm extends this base node with algorithm-specific data structures. In the EDF scheduler, the node is augmented with an embedded red-black-tree node used to link the corresponding thread into the EDF ready queue. It also contains an additional priority field that caches the minimum priority obtained from the priority aggregation tree.
Figure~\ref{fig:priority-levels} illustrates this structural relationship.

In RTEMS, job priority assignment, determined by its absolute deadline, is managed by the PR manager under the implicit-deadline assumption ($\forall \tau_i, D_i=T_i$) adopted in the default implementation. Notably, the implementation relies on legacy APIs prefixed with \textit{rate\_monotonic}, which were originally designed for fixed-priority scheduling, to handle job releases and deadline updates.

Following the workflow outlined in Figure~\ref{fig:priority-levels}, priority updates triggered by job releases or cancellations propagate through several layers before affecting dispatch decisions: The deadline is mapped to a priority in a priority node, the priority aggregation is updated, and the effective priority in the base node is potentially updated. If the effective priority changes, a pending update is recorded, denoting that the scheduler node's priority cache must be updated. When the scheduler node is updated, its position in the ready queue is adjusted. The schedule function then potentially leads to a new thread \emph{heir}, the thread that will be dispatched next, and signaling a deferred context switch. The PR manager calls both the release and cancel, as well as the potentially needed scheduler priority update functions.

Beyond the job release, cancellation, priority update and schedule functions, the EDF scheduler provides block, unblock and yield entry points which modify the ready queue and potentially trigger scheduling decisions.

These interactions between legacy priority-management infrastructure, scheduler-node updates, ready-queue manipulations and the internal mechanisms constitute the primary verification targets in this work. Both RTEMS 5 and 6 share the same overall architecture and workflow, with the main difference being in the internal mechanisms, with some of the EDF-specific dispatch logic being refactored into general uniprocessor scheduling helpers in RTEMS 6.

\subsection{Setup and Source Code Processing}
\label{sec:setup-and-source-code-processing}

The applied version of Frama-C includes machdep support for GCC extensions, accommodating compiler-specific constructs used in RTEMS, such as empty structs and flexible array members. We employ the corresponding RTEMS cross-compiler, \texttt{x86\_64-rtems5/6-gcc}, and select the \texttt{AMD64} Board Support Package (BSP),
the BSP available for \texttt{x86\_64} in RTEMS, 
supporting a uniprocessor configuration.

The EDF scheduler in RTEMS relies on the \texttt{tree.h} library from OpenBSD, which provides the red-black tree implementation used for ready queue management and priority aggregation within scheduler nodes. As shown  in Section~\ref{sec:key-assumptions}, 
we assume the correctness of this data structure by describing the behavior of its interface through contracts, abstracting its state as a set of nodes with associated predicates.

We verify the EDF Scheduler module, including its entry points, underlying mechanisms such as priority aggregation and ready-queue management, and the periodic release manager's release/cancel function. The gray-shaded region in Figure~\ref{fig:rtems-edf-flow} highlights the scope of the verification focused in this work. 
We provide abstract models for the ready queue and priority aggregation, state key invariants and helper lemmas, and annotate all relevant functions up to the RB-tree boundary. 
Verification is performed per entry point; contracts of internal functions are refined iteratively until all proof obligations are discharged.

\subsection{Verification Target}
\label{sec:4overview}
We outline the target of our verification, which encompasses the EDF scheduler module, underlying priority aggregation and ready queue mechanisms and the periodic release manager's release/cancel operations. Together, their contracts and their corresponding invariants establish the properties P1-P3. 

\begin{figure}[t]
    \centering
    \resizebox{\columnwidth}{!}{
        \begin{tikzpicture}
            \draw[draw,black, fill=yellow!15, dotted] (1.2,1) rectangle (8.8,-1.3);
            \draw[draw,black, fill=blue!15, dotted] (1.2,-1.3) rectangle (4.5,-2.9);
            \node[draw,align=left] (RM) at (-1.6,0.5) {PR Manager};
            \node[draw,dotted,align=left] (rtemsRate) at (-1.6,-0.8) {\footnotesize- Releasing Job \\\footnotesize - Cancelling Job};
            \draw[->, black] (-.5,-0.8) -- (0.8,-0.8);
            \draw[->, black] (-2,-1.3) |- (1.1,-2.1);
            \node[align=left] (rTrigger) at (-0.4,-1.9) {\footnotesize trigger priority update}; 
            \node[draw,align=left, fill=white] (EDF) at (2.85,0.6) {EDF Scheduler};
            \node[align=left] (edf1) at (2.85,-0.3) {\footnotesize\texttt{\_Release\_job} \\ \footnotesize\texttt{\_Cancel\_job}};
            \node[align=left] (edf2) at (2.85,-2.1) {\footnotesize\texttt{\_Update\_priority} \\
                \footnotesize\texttt{\_Schedule} \\
                \footnotesize\texttt{\_Yield} \\
                \footnotesize\texttt{\_Block/\_Unblock}    };
            \node[draw,align=left, fill=white] (THREAD) at (6.5,0.6) {Thread functions};
            \node[align=left] (t1) at (6.5,-0.5) {\footnotesize\texttt{\_add} \\
                \footnotesize\texttt{\_change} \\
                \footnotesize\texttt{\_remove}};
            
            \draw[black, dotted] (1.2,1) -- (1.2, -3.1);
            \draw[black, dotted] (4.5,1) -- (4.5, -3.1);
            \draw[black, dotted] (8.8,1) -- (8.8, -3.1);
            \draw[black, dotted] (-3,1) -- (-3, -3.1);
            
            \node[align=left, rotate=90] (priorityA) at (1,-0.5) {\scriptsize set deadline}; 
            \node[align=left, rotate=90] (priorityC) at (8.6,-0.5) {\scriptsize modify priorities}; 
            
            \draw[black, dotted] (-3.2,-1.3) -- (9, -1.3);
        \end{tikzpicture}
    }%
    \caption{Overview of verified EDF scheduler and thread priority functions, with prefixes omitted. Internal helpers are verified, but only entry points are shown. Yellow highlights deadline to priority propagation, while blue highlights core scheduling operations.}
    \label{fig:verification-map}
\end{figure}
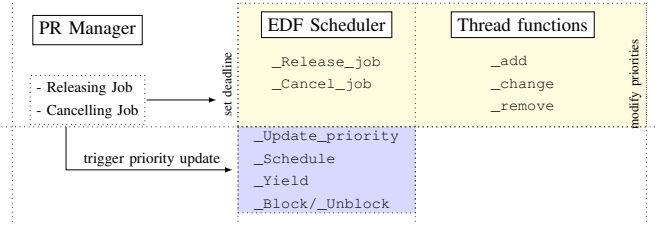

We categorize the relevant functions and invariants into two groups based on their roles in the EDF scheduling mechanism. The first group relates to the priority management and update pipeline (P1) as detailed in Section~\ref{sec:priority-manipulation-base}, while the second group focuses on the ready queue management and scheduling operations (P2-P3) given in Section~\ref{sec:4scheduling}. Figure~\ref{fig:verification-map} visually distinguishes these groups, using yellow for the first and blue for the second, though we note that update priority touches on all three properties.

We establish certain general structural properties as invariants, such as consistency of links between an EDF node and its thread, that are not directly related to the EDF-specific logic but are necessary for the overall correctness of the implementation. Beyond these, invariants are grouped by their role in the priority management pipeline or the scheduling operations, as detailed below.

\textbf{Deadline priorities~(P1):} In the PR Manager flow, the pathway for both job release and cancellation begins with either EDF's release or cancel function. When a job is released or canceled, it is necessary to modify the priority fields in its base scheduler node, ensuring that the priority aligns with the current absolute deadline. 

Based on an abstract model of the priority aggregation, we establish the following:

\begin{itemize}
	\item A postcondition on the release function ensures that the priority node's priority field is correctly set according to the job's absolute deadline (and, if not present, inserted into the aggregation tree). (P1)
	\item An invariant establishes that the priority in the base scheduler node is the minimum of all priority nodes in the aggregation tree. (P1)
	\item Another invariant ensures that the EDF scheduler node's priority field is consistent with the base node's priority or that an update priority call is pending to restore it. (P1)
\end{itemize}

These properties are verified for EDF's release, cancel entry points and the periodic release manager's release/cancel functions. For the periodic release manager, it is also ensured that the priority update is executed.

\textbf{Preemptive scheduling~(P2-P3):} For the scheduler, ready-queue updates and heir selection are triggered by thread state changes (block, unblock, yield, priority update) and explicit scheduling calls. The scheduler must ensure: the ready queue is correctly updated, and the thread heir is correctly set.

Based on an abstract model of the ready queue, we establish the following:

\begin{itemize}
	\item A postcondition on the unblock/block/yield functions ensures that the thread's EDF node is correctly added to or removed from the ready queue, and that the ready queue's state is updated accordingly as defined by the abstract model. (P2)
	\item An invariant that holds after every EDF entry point ensures that if the current heir is preemptible, it is the earliest ready job. (P3)
	\item Another invariant that holds after every EDF entry point, stating that if the current heir is not the executing thread, a dispatch is pending. (P3)
\end{itemize}

These properties are verified for the EDF update priority, unblock, block, yield and schedule entry points, as well as helper functions like initialization and update heir.

\subsection{Verification for Deadline-Driven Task Prioritization}
\label{sec:priority-manipulation-base}
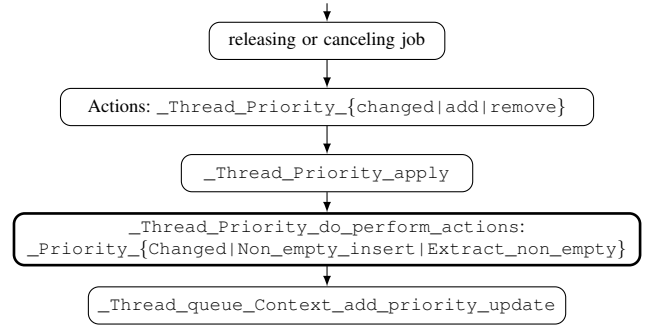
\begin{figure}[t]
	\centering
	\begin{tikzpicture}[scale=0.7, transform shape]
		\node [block, text width = 12em] (init) {releasing or canceling job};
		
		\node [block,  below of=init, node distance=1.25cm, text width = 28em] (tinsert) {Actions: \texttt{\_Thread\_Priority\_\{changed|add|remove\}}};
		
		\node [block,  below of=tinsert, node distance=1.25cm, text width = 15em] (apply) {\texttt{\_Thread\_Priority\_apply}};

		\node [block, line width=1pt, below of=apply, node distance=1.25cm, text width = 33em] (tactions) {\texttt{\_Thread\_Priority\_do\_perform\_actions}:\\ \texttt{\_Priority\_\{Changed|Non\_empty\_insert|Extract\_non\_empty\}}};

		\node [block,  below of=tactions, node distance=1.25cm, text width = 25em] (t2enqueue) {\texttt{\_Thread\_queue\_Context\_add\_priority\_update}};

		\draw[<-] ([yshift=0pt]init.north)  -- ++(0em,1em);
		
		\path [line] (init) -- (tinsert);
		
		\path [line] (tinsert) -- (apply);

		\draw[line] (apply) --  (tactions);

		\path [line] (tactions) -- (t2enqueue);
	\end{tikzpicture}
	\caption{Control flow of modifying the base scheduler nodes priority fields.}
	\label{fig:4controlflow}
\end{figure}

This subsection verifies Property~\ref{prop:deadline-priorities}. Figure~\ref{fig:4controlflow} provides an overview. All functions in this flow up to the RB-tree boundary have been annotated with ACSL contracts and verified.

In RTEMS, each thread (or job) contains a priority aggregation tree that represents all the priority entries and extracts the highest priority of the current job. The job release or cancellation process involves one of the following three \texttt{\_Thread} functions:
\begin{itemize}
	\item \texttt{\_Thread\_Priority\_changed} is used when the priority node is already part of the internal priority aggregation tree, necessitating an update to reflect the changes.
	\item \texttt{\_Thread\_Priority\_add} handles the addition of a new priority to the aggregation tree.
	\item \texttt{\_Thread\_Priority\_remove} is responsible for removing an existing priority from the tree.
\end{itemize}	

Afterwards, the \texttt{\_Thread\_Priority\_apply} function is called with an associated priority action type. Each action type influences the specific tree operation during the red-black tree's internal update.
It encapsulates both the priority node and the priority action type into a single object. This object is then utilized by the \texttt{\_Thread\-\_Priority\_do\_perform\_actions} function, %
resulting in one of two possible outcomes:
\begin{itemize}
	\item If the tree operation introduces a new minimum priority, it influences the task's positioning within the ready queue, so the scheduler node must be updated to reflect this new minimum value and an update action enqueued.
	\item If the tree operation does not result in a new minimum priority, no further action is required.%
\end{itemize}

After inserting the priority into the tree (if it exists), the  \texttt{\_Thread\_queue\_Context\_add\_priority\_update} function is called if a new minimum priority value emerges. The purpose is to enqueue the task into a scheduler queue for subsequent updating. In principle, this priority update could also be propagated to other priority aggregations due to e.g. synchronization-related priority adjustments. However, on the release/cancel path, the thread has to be either executing or ready, and this is therefore a noop.

The underlying priority combinators used for aggregation modification, \texttt{\_Priority\_\{Non\_empty\_insert, Changed, Extract\_non\_empty\}}, are each verified against the abstract aggregation model under the RB-tree interface assumptions, exhibiting the expected new-minimum or no-new-minimum behavior.

\subsection{Verification for Preemptive EDF Scheduling Operations}
	\label{sec:4scheduling}
	
	In this subsection, we detail the EDF scheduler operations establishing Property~\ref{prop:ready-queue} and Property~\ref{prop:scheduler-decisions}. The unblock, block, yield, schedule and update-priority entry points each update the ready queue and/or trigger an heir update. We focus on \texttt{\_Scheduler\_EDF\_Update\_priority} as the representative example. Following the previously described priority change from release or cancel, \texttt{\_Thread\_Priority\_update} checks for any pending priority updates, and if so, invokes the EDF update priority function for affected nodes. We have verified that the PR Manager's release and cancel functions correctly exhibit this behavior, updating both the base and EDF scheduler node priorities when needed.

\begin{figure}[t]
		\centering
	\scalebox{0.9}{
		\begin{tikzpicture}[scale=0.8, transform shape]
			\node [block, text width = 18em] (pending) {Pending update};

			\node [block, below of=pending, node distance=1.25cm, text width = 18em] (check) {\texttt{\_Thread\_Priority\_update}};
			\node [block, below of=check, node distance=1.25cm, text width = 18em] (init) {\texttt{\_Scheduler\_EDF\_Update\_priority}};
			\node [block,  below of=init, node distance=1.25cm, text width = 18em] (update) {Update cached priority, reinsert};
			\node [block,  below of=update, node distance=1.25cm, text width = 18em] (sched) {\texttt{\_Scheduler\_EDF\_Schedule}};
			\node [block,  below of=sched, node distance=1.25cm, text width = 18em] (heir) {Update heir};
			
			\draw[<-] ([yshift=0pt]pending.north)  -- ++(0em,1em);
			\draw[line] (pending) --  (check);
			\draw[line] (check) --  (init);
			\draw[line] (init) --  (update);
			\draw[line] (update) --  (sched);
			\path [line] (sched) -- (heir);
			
		\end{tikzpicture}
	}
		\caption{Control flow of executing priority update. Pending updates come from Fig.~\ref{fig:4controlflow}. Internal helpers for RTEMS 5/6 differ.}
		\label{fig:4controlflow2}
	\end{figure}
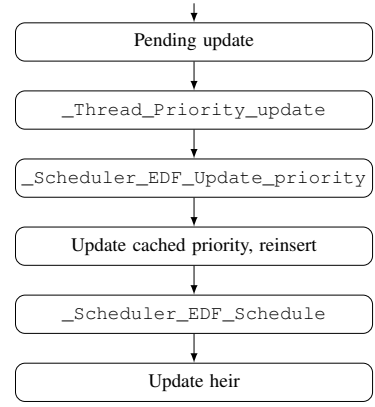

	This updates the cached priority field in the EDF scheduler node and adjusts the ready queue accordingly. If the new priority is earlier than the current heir's and the heir is preemptible, preparations for a context switch are initiated. This involves updating the thread heir, i.e., the pointer to the next processor owner scheduled for execution. This procedure is illustrated in Figure~\ref{fig:4controlflow2}. Postconditions include all the previously mentioned invariants and that the priority update is correctly reflected in the EDF scheduler node.

	\section{Verification of EDF Scheduling in FreeRTOS}\label{sec:freertos-edf}	
	To further evaluate the applicability of the proposed framework, we conduct a second case study on an EDF scheduler implementation in FreeRTOS. Unlike RTEMS, FreeRTOS provides a lightweight fixed-priority scheduling infrastructure by default, and EDF scheduling therefore requires kernel-level modifications. We describe the verification-oriented FreeRTOS setup, summarize the EDF-related scheduler changes, and explain their verification objectives.
	
	\subsection{Preliminary Setup}
	\label{sec:freertos-setup}
	Before describing the FreeRTOS scheduling infrastructure and our EDF modifications, we first clarify the FreeRTOS variant used in this case study. This setup is necessary because FreeRTOS is split into a portable kernel layer and a hardware-specific porting layer. Practical ports may contain platform-specific scheduler code or assembly routines that are outside the scope of Frama-C verification.
	
	The kernel layer provides the functionality needed by an RTOS to manage, schedule and execute tasks. The porting layer interfaces with a specific hardware platform, for example by providing the implementation of the periodic tick interrupt and context-switching routines. However, many ports such as the FreeRTOS port for the Espressif ESP32 microcontrollers~\cite{espidf} are highly optimized for the specific platform by implementing parts of the scheduler in assembly code and integrating platform-dependent code  into kernel-level scheduling paths. As Frama-C is not able to verify assembly code and modifications to the FreeRTOS kernel are necessary to provide an EDF scheduler, such official platform-specific distributions are not directly suitable for our verification purpose. 
	Hence, we modified a generic FreeRTOS V11.1.0 kernel to support EDF scheduling. As a hardware platform we use a Texas Instruments MSP430FR5994 microcontroller~\cite{msp430} and provide a corresponding FreeRTOS port. Our version keeps all scheduling-relevant operations in C, making the resulting EDF scheduler amenable to verification with Frama-C. We include our port of FreeRTOS for the MSP430FR5994 alongside our modifications to support EDF scheduling as part of the artifact~\cite{git_source_edf_verif}.

	\subsection{Scheduling Infrastructure}
	In general, FreeRTOS makes no distinction between jobs and tasks.
	Hence, there is only ever exactly one task instance that is either executing, ready, or blocked.
	By default, FreeRTOS uses static priorities that are assigned to each task at task creation.
	All information about a task is stored in its corresponding TCB object.
	The TCB however contains no information about task periods or deadlines.
	In general, FreeRTOS has no notion of explicit deadlines.
	A deadline in FreeRTOS is always assumed to be implicit and is only checked whenever a task has finished its execution.
	To realize periodic tasks, FreeRTOS uses the \texttt{xTaskDelayUntil()} function, where the task's period is supplied as a function parameter.
	This function first determines the next release time of the task.
	Then, if the next release time has not already passed, the task's deadline is not violated and the task is inserted into the global delayed list, sorted by its next release time.
	Finally, the task yields, i.e., the scheduler is invoked to context switch to the next highest priority task.
	If the task has missed its deadline, the task stays in the ready queue. 

	For the scheduling behavior considered in this work, the relevant task states are represented primarily by ready lists and delayed lists.
	Tasks that are ready to be executed are stored in the ready list and tasks that are blocked until a later time unit are stored in a global delayed list\footnote{FreeRTOS also maintains additional lists, such as overflow delayed lists for tick-count wraparound. These mechanisms are part of the original kernel infrastructure and are not specific to our EDF modification.}. 
	In standard FreeRTOS, the ready list is kept sorted by the static priorities of the tasks, while the delayed list is sorted by the next release time of any task contained therein.

	A common FreeRTOS idiom for implementing periodic tasks is to use \texttt{xTaskDelayUntil()} together with the periodic tick interrupt, which iterates through the global delayed list and moves any task that becomes ready at the current tick count to the ready list, and performs a scheduling decision.
	The fixed-priority scheduler always selects the head element in the global ready list as the currently running task.
	The currently executing task is denoted by a global variable \texttt{pxCurrentTCB} that points to the task's TCB.
	At any point outside of the scheduler, the currently executing task is pointing to the head element of the ready list.
	Whenever a task gets released to the ready list, the scheduler determines if the currently running task is still the head element.
	If that is no longer the case, a context switch to the task at the head of the ready list is performed.

	\subsection{EDF Scheduling in FreeRTOS}
	A straightforward way to retrofit EDF scheduling into FreeRTOS is to preserve the existing list-based scheduling architecture and replace static priorities by absolute deadlines as the scheduling key. 
	This requires modifications to three central parts of the kernel: the task metadata stored in the TCB, the update of timing information in \texttt{xTaskDelayUntil()}, and the ready-list ordering used by the scheduler.	
	First, the current absolute deadline of a task must be stored.
	For this, we replace the static priority field of the TCB by a current absolute deadline field \texttt{xDeadline}.
	Consequently, when a task is created, the value supplied in place of the original static priority denotes the task's initial absolute deadline.

	Second, the \texttt{xDeadline} must be updated before the next time that the task will be released. Since we do not add any additional fields to the TCB to keep its original size, we can only rely on one value that stores the current absolute deadline. Therefore, we integrate the setting of the new deadline into the \texttt{xTaskDelayUntil()} function to exploit the period parameter as the relative deadline. As a result, we force implicit deadlines.

	Third, the \texttt{xDeadline} field must be respected by the scheduler whenever it is invoked.
	That means, when a task is inserted to the ready list, the \texttt{xDeadline} is used as the sorting criterion.
	For a sorted insertion into a list, FreeRTOS uses an internal list item field in the TCB.
	This field usually contains the next release time of the task, as a sorted insertion by default only occurs when a task is added to the delayed list.
	We only replace the list item value by the \texttt{xDeadline} prior to the insertion to the ready list, such that the ready list is kept sorted after the current absolute deadlines of the tasks.

	Besides these modifications, all internal functions where the scheduler had previously used the static task priorities for comparing which task to execute are modified to take the current absolute deadlines into account.
	The overall architecture of the FreeRTOS scheduler remains unchanged.

	\subsection{Verification Target}
	We outline our verification target for the EDF scheduler in FreeRTOS.
	As described in Section~\ref{sec:freertos-setup}, our FreeRTOS variant keeps all scheduling-relevant operations in C. 
	We therefore cover the tick increment, delay, suspend, resume, and task-switch selection functions, together with their underlying helper functions. 
	Together, the verified contracts and invariants establish Properties~\ref{prop:deadline-priorities}--\ref{prop:scheduler-decisions}.

	As in the RTEMS case study (Section~\ref{sec:key-assumptions}), we treat the list module as a correct component. Similarly, we also use Frama-C's \texttt{Volatile} option to reason about the dispatch flag. However, in addition to the dispatch flag, several other global variables such as the current TCB pointer and the delayed list are also volatile in FreeRTOS, requiring us to operate under the same scheme for them.

	\textbf{Deadline priorities (P1)}: As no priority aggregation mechanism is used, P1 involves only two aspects: 1) the delay function correctly sets the deadline and 2) maintaining an invariant that ensures that the deadline field in the TCB corresponds to the value in the ready list item field used for sorting. This invariant is maintained across all entry points.

	\textbf{Ready queue management (P2)}: We establish invariants that ensure the ready, delay and suspend are well-formed and disjoint lists, and that the ready and delay list have their head pointing to the minimum element. We combine these invariants into a single scheduler context invariant that is maintained across all entry points.
	The lists are correctly updated in the delay, suspend and resume functions, with e.g. all jobs that become ready during a tick being moved from the delayed list to the ready list. To ensure this we use loop invariants.
	
	\textbf{Scheduler decisions (P3)}: The task switch function correctly selects the task with the earliest deadline and updates \texttt{pxCurrentTCB} accordingly when the scheduler is running or sets a dispatch pending flag. An invariant is maintained across all entry points that the current TCB has the smallest deadline among all ready tasks (or that the dispatch flag is set). In the tick ISR, the task switch is immediately triggered after the tick increment call, ensuring that afterwards the current TCB is always the task with the earliest deadline when the scheduler is not suspended.	
	
	\textbf{Finding.} Updating the deadline in \texttt{xTaskDelayUntil()} is natural because the function computes the next release time and receives the period parameter. Our verification, however, revealed an overrun case: If a task calls \texttt{xTaskDelayUntil()} after its next release time has already passed, FreeRTOS keeps the task in the ready list instead of delaying it. Under fixed-priority scheduling this is harmless. Under EDF, however, updating \texttt{xDeadline} while the task remains in the ready list may invalidate the ready-list ordering, thereby violating P1 and P2. The verified fix is to reinsert the task into the ready list whenever an overrun causes the task to remain ready after its deadline update.

	\section{Results and Discussions}

The proposed framework was applied to multiple EDF scheduler realizations, including RTEMS 5, RTEMS 6, and FreeRTOS. Across these implementations, the verification focused on the three core EDF correctness properties defined in Section~\ref{sec:rtems-edf-properties}: deadline-driven task prioritization, ready queue management, and preemptive scheduling behavior. The verification sources and the setup for reproducing the evaluation tables are available in our artifact~\cite{git_source_edf_verif}.

\subsection{Verification Results}
	For RTEMS and FreeRTOS, the verification confirms that the EDF scheduler
	implementations satisfy Properties~\ref{prop:deadline-priorities}-\ref{prop:scheduler-decisions} under the assumptions of
	Section~\ref{sec:key-assumptions}.
	\begin{itemize}
	  \item \textbf{P\ref{prop:deadline-priorities}:} In RTEMS, the priority
	    pipeline maps every deadline change at job release/cancel into the minimum of the aggregation tree and propagates it into the EDF scheduler node. In FreeRTOS, \texttt{xTaskDelayUntil} refreshes \texttt{xDeadline} at
	    every release, and an invariant ties the ready-list sort key to it.
	
	  \item \textbf{P\ref{prop:ready-queue}:} In RTEMS, all ready-queue mutations preserve the ready-queue model with the earliest-deadline job at the head and correctly insert and remove jobs. In FreeRTOS, the context invariant ensuring valid ready, delayed and suspended lists, is preserved by every verified entry point and correctly moves jobs between the lists.
	
	  \item \textbf{P\ref{prop:scheduler-decisions}:} In RTEMS,
	    schedule picks the earliest-deadline ready job as heir and requests a context switch when the running thread is preemptible. In FreeRTOS, the tick ISR ensures the EDF property holds unless the scheduler is suspended, while other functions preserve the EDF property or request a dispatch. 
	\end{itemize}

\begin{table}[tp]
    \footnotesize
    \centering
    \caption{Derived proof goals and required time of RTEMS verification (each entry: \emph{RTEMS 5 / RTEMS 6}).}
    \label{tab:rtems-stats}
\begin{tabular}{>{\raggedright}p{0.33\linewidth} >{\centering}p{0.05\linewidth}>{\centering}p{0.2\linewidth}>{\centering\arraybackslash}p{0.2\linewidth}}
    & \multicolumn{2}{c}{\textbf{Proof Goals}} & \multirow{2}{*}{\textbf{Approx. Time}}\\ \cmidrule{2-3}
	& Qed & Alt-Ergo & \\ \hline \midrule
    \texttt{Initialize} & 10/10 & 5/5 & {$ 4.9/5.0\,s$} \\
    \texttt{Node\_initialize} & 15/15 & 0/0 & {$ 0.8/0.8\,s$} \\
    \texttt{Block} & 41/42 & 19/14 & {$ 8.6/5.1\,s$} \\
    \texttt{Schedule} & 13/19 & 2/3 & {$ 3.6/3.9\,s$} \\
    \texttt{Yield} & 30/35 & 10/10 & {$ 4.3/4.5\,s$} \\
    \texttt{Unblock} & 67/81 & 21/17 & {$ 10.8/12.6\,s$} \\
    \texttt{Update\_priority} & 139/166 & 64/62 & {$ 23.5/35.0\,s$} \\
    \texttt{Release\_job} & 144/148 & 133/130 & {$ 51.6/51.7\,s$} \\
    \texttt{Cancel\_job} & 157/156 & 82/81 & {$ 30.1/29.9\,s$} \\
    \texttt{Thread\_Priority\_\{*\}} & 119/189 & 14/55 & {$ 84.6/15.2\,s$} \\
    \texttt{RM\_Release\_job} & 67/69 & 39/38 & {$ 30.9/33.4\,s$} \\
    \texttt{RM\_Cancel} & 69/73 & 39/39 & {$ 23.2/27.3\,s$} \\
    \texttt{Scheduler helpers} & 88/84 & 12/3 & {$ 28.9/26.0\,s$} \\
    \texttt{Priority helpers} & 625/635 & 339/337 & {$ 401.6/152.1\,s$} \\
    \midrule
    \textbf{Total} & \textbf{1584/1722} & \textbf{779/794} & {$ 707.5/402.6\,s$} \\
\end{tabular}
\end{table}
\begin{table}[tp]
	\footnotesize
	\centering
	\caption{Derived proof goals and required time of FreeRTOS verification.}
	\label{tab:freertos-stats}
\begin{tabular}{>{\raggedright}p{0.3\linewidth} >{\centering}p{0.05\linewidth}>{\centering}p{0.2\linewidth}>{\centering\arraybackslash}p{0.2\linewidth}}
    & \multicolumn{2}{c}{\textbf{Proof Goals}} & \multirow{2}{*}{\textbf{Approx. Time}}\\ \cmidrule{2-3}
    & Qed & Alt-Ergo & \\ \hline \midrule    
    \texttt{vTaskSwitchContext} & 32 & 2 & {$ 1.1\,s$} \\
    \texttt{vTaskSuspend} & 45 & 25 & {$ 11.9\,s$} \\
    \texttt{vTaskResume} & 40 & 21 & {$ 4.8\,s$} \\
    \texttt{xTaskDelayUntil} & 60 & 44 & {$ 44.6\,s$} \\
    \texttt{xTaskIncrementTick} & 98 & 43 & {$ 44.7\,s$} \\
    \texttt{Task helpers} & 154 & 21 & {$ 14.3\,s$} \\
    \midrule
    \textbf{Total} & \textbf{429} & \textbf{156} & {$ 121.4\,s$} \\\end{tabular}
\end{table}

	\subsection{Verification Effort and Reusability}
	The framework carried across all three case studies at the structural
	level: the three correctness properties and invariants maintaining them, the abstract-data-structure boundary, and specific strategies like the ghost-mirroring discipline for volatile scheduler flags applied without change. Within the RTEMS family this structural reuse extended to many contracts, as RTEMS~5 and RTEMS~6 share the same general scheduler architecture, with their differences mainly in an internal refactoring and internal helpers. 
	FreeRTOS applies the methodology on a substantially different infrastructure, with several doubly linked lists and no clear API boundary between the scheduler and the underlying data structure manipulation. The same methodology and overall strategies for approaching the correctness properties were applicable, but more proof engineering was required to find fitting data structure invariants and contracts to work around these differences.
	
	The derived proof goals for both built-in Qed and selected proofs to Alt-Ergo, and required time of all covered parts of the verification of RTEMS 6, RTEMS 5 and FreeRTOS, via Frama-C and \texttt{wp}, are presented in Table~\ref{tab:rtems-stats} and Table~\ref{tab:freertos-stats}, respectively.
	The process was executed with 24 threads on an AMD Ryzen 9 3900X with 64 GB main memory.

	\subsection{Practical Considerations for Source-Level Verification}
	\label{sec:limitations}
	
	Applying deductive verification directly to RTOS source code requires bridging the gap between solver-friendly specifications and low-level implementation idioms. 
	In our case studies, the main challenges arise from pointer-based data structures, unsafe casts, and macro-heavy kernel code. 
	They are handled through explicit verification boundaries, auxiliary contract clauses, and proof-oriented wrappers where needed.
	
	First, ACSL can express separation between memory regions, but proving separation properties over larger data structures such as multiple linked lists is often difficult for the solver. 
	For instance, in FreeRTOS, modifying the delayed list should preserve predicates over the ready list. 
	Although this can be specified by separating the reachable elements of the two lists, the solver does not always establish automatically that such separation preserves list-level properties across list manipulations. 
	We thus state preservation conditions explicitly in the contracts of list operations, requiring properties of untouched lists to remain unchanged. In the same way, in RTEMS we annotate internal helpers with explicit preservation postconditions, leading to at times verbose contracts and more manual proof effort. For the RM functions, it was necessary to mark the update queue as a static global, as Frama-C was not able to verify that deallocating the empty update queue from the stack did not affect the ready queue. This does not alter the behavior in the non-SMP case.

	Second, system-level C code often uses casts to implement low-level abstractions and structural reuse. As discussed in Section~\ref{sec:framac-and-mm}, the selected Frama-C memory model supports such casts, but does not itself prove that every cast preserves the intended structural interpretation. Accordingly, our verification treats the correctness of these representation assumptions as part of the trusted boundary of the source code.

	Finally, FreeRTOS makes extensive use of macros for list operations. Since Frama-C verifies preprocessed C code and ACSL contracts cannot be attached directly to macros, we introduce proof-only wrapper functions for the relevant macro calls. These wrappers provide explicit contracts while preserving the original implementation semantics. The required source-level adaptation is therefore minimal and limited to making existing operations visible to the verifier.

	\section{Related Work}
	\label{sec:related-work}
	
Providing a verified RTOS can be approached in two main ways. 
The first approach is to develop a new RTOS, or key RTOS components, together with formal correctness proofs. 
For complete operating-system kernels, representative efforts include the verification of the seL4 microkernel in Isabelle/HOL~\cite{klein2009sel4}, which required parts of the microkernel to be manually reimplemented for verification. 
Gu et al.~\cite{DBLP:conf/osdi/GuSCWKSC16} presented CertiKOS, a framework for building concurrent operating-system kernels whose layers are verified in Coq~\cite{DBLP:series/txtcs/BertotC04}. 
Guo et al.~\cite{DBLP:conf/cav/GuoLLRS19, guo2021generic} further connected verified schedulability analysis with CertiKOS using the Proven Schedulability Analysis (PROSA) framework.

Scheduler-specific verification has also been studied in this line of work.  
Vanhems et al.~\cite{DBLP:conf/rtas/VanhemsRNG22} proposed a formal proof methodology for an EDF scheduler. They first prove the correctness of the election function in Coq and then lift this proof to the implementation level through refinement steps into a shallow embedding of a subset of C, which is then translated to C using the Digger tool, running on top of the Pip protokernel~\cite{DBLP:journals/eceasst/JomaaTNGH18}. These approaches provide strong correctness guarantees, but they usually require the scheduler or kernel to be developed within a verification-oriented framework. There is currently no way to use these methods to verify existing schedulers or new schedulers implemented in existing RTOS kernels. Given the widespread use of existing RTOS implementations, this therefore presents a significant gap in the state of the art for applying these verification techniques to real-world systems.

The second approach focuses on existing RTOS implementations, which is the focus of this work.
Model checking has been used to analyze RTEMS, Trampoline/OSEK, ARINC 653 systems, partitioned RTOS models, Contiki, and FreeRTOS scheduler/interrupt behavior~\cite{DBLP:conf/icfem/GadiaAB16, DBLP:conf/rtns/BoukirBD20, DBLP:conf/sac/SinghDE21, staroletov2020formal, DBLP:journals/tcps/MousaviEM23, DBLP:conf/icfem/LinW24}.
Specifically, Gadia et al.~\cite{DBLP:conf/icfem/GadiaAB16} modeled RTEMS code in Java and verified it using Java Pathfinder. 
Lin and Wang~\cite{DBLP:conf/icfem/LinW24} analyzed the FreeRTOS scheduler on ARM Cortex-M4 cores by modeling both the scheduler and the relevant interrupt mechanisms. 
These approaches are effective for finding subtle design errors, but may suffer from state-space explosion and often verify translated or abstracted models rather than the source implementation directly~\cite{DBLP:conf/icfem/GadiaAB16,baier2008principles}. 

To reduce the gap between the verified artifact and the implementation, deductive verification~\cite{DBLP:journals/cacm/Hoare69,floyd_flowcharts_1967} can be used to verify source code directly against formal specifications. 
This approach is supported by tools such as VerCors~\cite{DBLP:conf/fm/BlomH14}, VCC~\cite{DBLP:conf/tphol/CohenDHLMSST09} and Frama-C~\cite{cuoq2012frama}. 
VCC has been used to verify parts of the Microsoft Hyper-V hypervisor~\cite{10.1007/978-3-642-05089-3_51}, while Frama-C has been applied to safety-critical embedded software such as the Paparazzi UAV autopilot~\cite{pollien2021verifying}. 
In the RTOS context, Chong and Jacobs~\cite{chong2021formally} formally verified the interprocess communication mechanism of FreeRTOS. 
Tasche et al.~\cite{DBLP:journals/tecs/TascheHH25} proposed a deductive-verification approach for cooperative FreeRTOS applications using VerCors, combining an encoding of FreeRTOS semantics with automatically generated reachable abstract-state invariants.
Shi et al.~\cite{DBLP:journals/tcad/ShiECC22} introduced a Frama-C-based framework for resource synchronization protocols, studying the implementations in RTEMS. 
These works demonstrate the applicability of deductive verification to RTOS-related software, but they do not target the verification of EDF scheduler implementations in existing RTOS kernels.

Other verification approaches address related but distinct aspects of operating-system correctness. 
Nicole et al.~\cite{DBLP:conf/rtas/NicoleLBR21} developed a fully automated verification process for small OS kernels at the binary level, ensuring the absence of privilege escalation and runtime errors. 
However, this approach does not verify whether the implementation satisfies high-level functional properties derived from the scheduler design. 
Some formal methods focus on timing or schedulability analysis rather than implementation correctness. 
For instance, Bozhko et al.~\cite{DBLP:conf/ecrts/BozhkoB20} established a formal foundation for the busy-window principle through an abstract response-time analysis independent of specific scheduling policies and workload models, with all definitions and proofs mechanized in the Coq proof assistant.

	\section{Conclusion} 
	\label{sec:conclusion}
	
	In this work, we proposed a general framework for verifying the correctness of EDF scheduler implementations in existing RTOSes. We formally defined three correctness properties that all EDF schedulers must satisfy. These properties capture the fact that EDF correctness is not limited to selecting the earliest-deadline job, but also depends on the consistent propagation and maintenance of deadline-derived priorities across scheduler data structures and task release operations. 
	
	We applied the framework to the uniprocessor EDF scheduler in RTEMS 5, RTEMS 6 and FreeRTOS, performing the first deductive verification
	under explicitly stated assumptions on kernel infrastructure and data-structure operations. We release our verification effort as an artifact~\cite{git_source_edf_verif}.

	The FreeRTOS case study further demonstrates the usefulness of the framework by exposing an overrun-related edge case in a straightforward EDF retrofit, where updating a task's deadline without reinserting it into the ready list may violate the ready-queue ordering.
	The results show that our framework can be applied to EDF scheduler logic in legacy or repurposed RTOS infrastructures without re-engineering the entire kernel.

\label{last-page}

\section*{Acknowledgments}

This result is part of a project (PropRT) that has received funding from the European Research Council (ERC) under the European Union’s Horizon 2020 research and innovation programme (grant agreement No. 865170). This work has received funding from the DFG Priority Program ``Disruptive Memory Technologies'' (SPP 2377) as part of the project ``ARTS-NVM'' (502308721). It is further supported by the DFG Project ``One-Memory'' (405422836).

	\bibliographystyle{abbrv}

\bibliography{arxiv-real-time}

\end{document}